\documentstyle[12pt]{article}
\def\beq{\begin{equation}}
\def\eeq{\end{equation}}
\def\bea{\begin{eqnarray}}
\def\eea{\end{eqnarray}}

\def\ba{\begin{array}}
\def\ea{\end{array}}
\begin{document}
\begin{titlepage}
\rightline{SNUTP-97-130}
\rightline{UM-TG-201}
\rightline{hep-th/9709096}
\def\today{\ifcase\month\or
        January\or February\or March\or April\or May\or June\or
        July\or August\or September\or October\or November\or December\fi,
  \number\year}
\vskip 1cm
\centerline{\Large \bf M Theory Fivebrane Interpretation for Strong }
\centerline{\Large \bf Coupling Dynamics of $SO(N_c)$ Gauge Theories}
\vskip 1cm
\centerline{\sc Changhyun Ahn$^{a,}$\footnote{ chahn@spin.snu.ac.kr},  
Kyungho Oh$^{b,}$\footnote{ oh@arch.umsl.edu}
and Radu Tatar$^{c,}$\footnote{tatar@phyvax.ir.miami.edu}}
\vskip 1cm
\centerline{{\it $^a$ Dept. of Physics, Seoul National University,
Seoul 151-742, Korea}}
\centerline{{ \it $^b$ Dept. of Mathematics,
 University of Missouri-St. Louis,
 St. Louis, Missouri 63121, USA}}
\centerline{{\it $^c$ Dept. of Physics, University of Miami,
Coral Gables, Florida 33146, USA}}
 \vskip 3cm
\centerline{\sc Abstract}
\vskip 0.2in

We study M theory fivebranes to understand the moduli space of vacua of 
$N=1$ supersymmetric $SO(N_c)$ gauge theory with $N_f$ flavors in four 
dimensions. We discuss how the various branches of this theory arise in the 
string/M
theory brane configurations and compare our results with the ones
obtained  earlier by Intriligator and Seiberg in the context of field theory. 
In the M theory approach, we explain the various branches from 
the asymptotic position of semi-infinite
D4 branes in the $w=x^8+i x^9$ direction, which is closely related
 to the eigenvalues of the 
meson matrix $M^{ij}=
Q^{i}_a Q^{j}_a$ where $Q^{i}_a$ is a squark multiplet( $i=1, \cdots, 2N_f$ 
and $
a=1, \cdots, N_c$).
In M theory, these branches are explained by observing a new phenomena 
which did not occur for the gauge groups
$SU(N_c)$ or $Sp(N_c)$.

\vskip 0.6in
\leftline{September, 1997}
\end{titlepage}
\newpage
\section{Introduction}
In the last year much progress has been made in deriving field theory
results from string/M theory. Starting with the work of Hanany and Witten
for the case of 8 supercharges in three dimensions \cite{hw}(see also
\cite{bo1}) and its generalization to test the $N=1$ field theory
dualities in four dimensions \cite{se} in the paper of
Elitzur, Giveon and Kutasov \cite{egk},
many important results were obtained, and it turned out to reveal new aspects of both
field theory and string theory. 

For the $N=1$ theories two approaches were developed, one based on the
work of \cite{egk,eva} and the other based on
wrapping D6 branes around three cycles in  Calabi-Yau manifolds 
which was initiated in \cite{ov} and
generalized in \cite{ao}. In all these cases, the theory is in the
type IIA set-up with weak string and gauge couplings, and the branes are rigid.

A natural way to go to strong coupling in string theory is to promote
the M theory set-up. The M theory approach was pioneered in the seminal
work of Witten \cite{w1}. A simple observation  that both
D4 branes and NS5 branes of  type IIA setup can be interpreted as
 a single fivebrane of
M theory,  gives a unified description. The world volume of the fivebrane
can be described as a product space of
four dimensional spacetime and  the Seiberg-Witten curve \cite{sw}
for the $SU$ gauge group. The result was generalized to
$SO$ and $Sp$ groups \cite{lll,bsty1}, and many other very interesting results
have been obtained 
\cite{ah,ba,w2,hk,nos,hy,noyy,ss,ll,lyk,bo,mi,hoo,eva1,aot1,bisty1}.

In the present work, we continue to follow the lines initiated in \cite{w2,hoo} and
generalized in \cite{bo,aot1} 
which exploits the fivebrane of M theory in order to describe strong coupling
dynamics of $N=1$ theories. We generalize to the case of the
$SO(N_{c})$ gauge group. 
As is known from field theory, for the $SO$ group there is
a clear distinction between the Higgs phase and the confining phase, 
and furthermore,
in different phases, the effective superpotential takes different forms.  
It is well known \cite{se,is} that, for $N_{f}\le N_{c}-5$, in quantum theory
there exists a superpotential and the theory does not have a ground state. 
For $N_{f}=N_{c}-4$ and $N_{c}-3$, the low energy theory has two branches, one
with a superpotential and no ground state, the other without superpotential
where there exist ground states. For $N_{f}=N_{c}-2$, the
unbroken group is $SO(2)$ and the theory is abelian, and moreover there are vacua with
monopole and dyon condensations. For $N_{f}\ge N_{c}-1$, there  
exists a non-abelian magnetic dual theory with the dual gauge group $SO(N_{f}-N_{c}-4)$. 
We find an agreement with field theory results except in some subtle points in which
we could not discuss the massless matter which appear at the vacua and 
in the dual theory (Note that
similar problems were encountered in \cite{aps} where the $N=1$ dualities were
derived by breaking $N=2$ supersymmetry to $N=1$). 

In section 2, we study the moduli space of $N=1$ theory obtained from the $N=2$
theory by adding a mass term of the adjoint chiral multiplet. In section 3,
we discuss the Higgs branches of the $N=2$ theory and the resolution of singularities.  
In section 4,
we rotate the fivebrane to break the supersymmetry to $N=1$ for a
non-zero value of the adjoint mass and we take the mass of the
adjoint to infinity and compare the deformation space of the brane configuration
with the moduli space of vacua of the $N=1$ theory. Finally in section 5, 
we will conclude and point out the relevant future directions.  

\section{Field theory results}
\setcounter{equation}{0}

We give a brief summary of the field theory results by following the
same arguments done in \cite{hoo,aot1}. 

\subsection{From $N=2$ theory to $N=1$ theory}

Along the line of \cite{aps,hms}, we start from $N=2$ supersymmetric
$SO(N_{c})$ theory with $2N_{f}$ squark multiplets $Q^{i}_{a}$ where
$i = 1, \cdots, 2N_{f}$ and the color index $a = 1, \cdots, N_{c}$. 
The superpotential for the massless quarks is
\beq
\label{superpot1}
W = \sqrt{2}Q_{a}^{i}\Phi_{ab}Q^{j}_{b}J_{ij}
\eeq
where $\Phi_{ab}$ is a chiral multiplet in the adjoint representation 
of the gauge group, and $J$ is the symplectic
$2N_{f}\times 2N_{f}$ dimensional metric used to raise or lower flavor
indices. 
We want to break $N=2$ supersymmetry to $N=1$ by inserting a mass term of the
adjoint field $\Phi_{ab}$. The superpotential (\ref{superpot1})
then modifies to
\beq
\label{superpot2}
 W = \sqrt{2}Q_{a}^{i}\Phi_{ab}Q_{b}^{j} + \mu \mbox{Tr}(\Phi^{2}).
\eeq
We can integrate out the $\Phi_{ab}$  by using its F-term equation:
\beq
Q^{i}_{a}J_{ij}Q^{j}_{b}-\sqrt{2}\mu\Phi_{ab} = 0.
\label{fterm}
\eeq
As the mass $\mu$ is increased beyond the scale of the asymptotic free
$N=2$ theory, and we integrate out the adjoint field, by one loop
matching condition between the $N=1$ and $N=2$ theory,  we get the dynamical 
scale $\Lambda_{N=1}$ given by:
\beq
\Lambda_{N=1}^{3(N_{c}-2)-2N_{f}} = \mu^{N_{c}-2}\Lambda_{N=2}^{2(N_{c}-2)-2N_{f}}.
\label{scale1}
\eeq
Plugging (\ref{fterm}) into (\ref{superpot2}) we can write down the expression of the
superpotential only in terms of $M^{ij} = Q^{i}_{a}Q^{j}_{a}$. 
The final form of the superpotential is given by:
\beq
\Delta W = \frac{1}{2\mu} \mbox{Tr}(MJMJ).
\label{superpot3}
\eeq
The system below the energy scale $\mu$ can be considered as $N=1$ theory
with the tree level superpotential (\ref{superpot3}) and with the dynamical
scale $\Lambda_{N=1}$ given by (\ref{scale1}). 
In the next two subsections we will discuss  the gauge group $SO(2N_{c})$
and  $SO(2N_{c}+1)$ for various values of $N_f$ and $N_c$.

\subsection{$SO(2N_{c})$ groups}

Our discussion is based on the field theory results obtained
 earlier in \cite{se,is}.
Since the number of flavors is $2N_{f}$, the difference
between the number of flavors and the number of colors is always even.

$\bullet \;\;\;0 \le 2N_{f} \le 2N_{c}-6$

A supepotential is dynamically
generated by strong coupling effects \cite{ads}. For a generic value of
$2N_{f}$, the ADS superpotential is given by 
\beq
W_{ADS} = (N_{c}-N_{f}-1)\omega_{2N_{c}-2N_{f}-2}(\frac{16\Lambda_{N=1}^{6N_{c}-
2N_{f}-6}}{\mbox{det} M})^{\frac{1}{2N_{c}-2N_{f}-2}}  
\label{ads}
\eeq
where $\omega_{2N_c-2N_f-2}$ denotes $(2N_c-2N_f-2)$-th root of unity.
For large but finite $\mu$, at the scale far below $\mu$
but larger than $\Lambda_{N=1}$, the superpotential (\ref{superpot3})
can be regarded as a perturbation of the ordinary $N=1$ theory.
Thus the total
effective superpotential is
\beq
W_{eff} = W_{ADS} + \Delta W.
\label{eff}
\eeq
As explained in \cite{hoo,aot1}, this form for $W_{eff}$ is exact for any non-zero
values of $\mu$ by holomorphy and symmetry argument. 
Because the meson field $M$ is a symmetric matrix it can be brought to diagonal form 
after an appropriate similarity
transformation, where the diagonal elements are given by
$(m_{1},\cdots, m_{N_{f}}, m_{1}, \cdots,
m_{N_{f}})$.
By extrematizing $W_{eff}$ with this explicit form of $M$,
 all the
diagonal elements of $M$ become equal and are given by:
\beq
\label{moduli}
m = 2^{\frac{2}{2N_{c}-2-N_{f}}}\mu\Lambda_{N=2}.
\eeq   
The value of $m$ in (\ref{moduli}) describes the moduli space of the $N=1$ 
theory in the presence of the perturbation to the ADS superpotential. 
In the limit of $\mu \rightarrow \infty$ keeping $\Lambda_{N=1}$ finite, 
 $m$ diverges so that $\Delta W$ goes to 0. Thus there is
no supersymmetric vacua. 

$\bullet \;\;\;2N_{f} = 2N_{c}-4$

As explained already in \cite{is}, there exist two branches of the theory.
The origin of these branches is due to the fact that SO(4) is isomorphic to
the product $SU(2)_{L}\times SU(2)_{R}$ and the gaugino condensation in the
product gauge group generates the superpotential:
\beq
\label{ads1}
W_{ADS} = \frac{1}{2}(\epsilon_{L}+\epsilon_{R})
\left(\frac{16\Lambda_{N=1}^{2(2N_{c}-1)}}
{\mbox{det}M}\right)^{1/2},
\eeq
where $\epsilon_{L,R}$ are $\pm 1$. The case of $\epsilon_{L}\times
\epsilon_{R}=1$ gives the first branch while the case of
$\epsilon_{L}\times\epsilon_{R}=-1$ does the second one.
The first branch is the continuation of
the (\ref{eff}) to the case $2N_{f} = 2N_{c}-4$. 
By taking the extremum with
respect to $M$, which has diagonal form, 
we obtain again $m =(4(N_c-2))^{\frac{1}{N_c}} \mu \Lambda_{N=2}$ which will lead to the
same limit with no vacuum as previous case when $\mu$ goes to infinity.
However, the second branch is not a continuation of (\ref{eff}).
For this case, $W_{ADS}=0$ and the only thing that is to be extrematized is 
$\Delta W$ which just gives $m=0$. So there is a  vacuum at the
origin, i.e. $M=0$. This is in complete agreement with the theory \cite{is} without the
perturbation of $\Delta W$. So quantum mechanically, at the origin only the
$M$ quanta are massless and this describes the confinement of the elementary degrees of
freedom. The global chiral symmetry is unbroken, so we have confinement 
without chiral symmetry breaking, a new feature which did not appear in the case
of $SU$ gauge group with $N_{f}=N_{c}$. 

$\bullet \;\;\;2N_{f}=2N_{c}-2$

There is no superpotential
because in (\ref{ads}) the inverse power of the right hand side vanishes, $2N_{c}-2N_{f}-2=0$ 
and the theory has a quantum moduli
space of vacua given by the expectation values of $M^{ij}$. The gauge group 
$SO(2N_{c})$ is broken to $SO(2)$ and the theory is in a Coulomb phase with
a massless supermultiplet. 
As a continuation of the previous case, there exist two branches, one at
$m \sim \mu\Lambda_{N=2}$, the other  at $m=0$. Using now the RG matching
relation (\ref{scale1}),  
$\Lambda_{N=1}^{4N_{c}-4}=\mu^{2N_{c}-2}\Lambda_{N=2}^{2N_{c}-2}$,
we get $\mbox{det} M \sim \Lambda_{N=1}^{2N_{c}-4}$ which is exactly the
same as $U_{1}$ obtained in \cite{is}. 
The other vacua is at $\mbox{det} M=0$
which is again in accordance with $U=0$. At this stage, let us stress that we
could not obtain any information about the monopoles or dyon condensation, but
we could check the existence of the vacua of the theory.

$\bullet \;\;\;2N_{f}\ge 2N_{c}$

In  this case, very interesting phenomena have been
observed in \cite{is,se}.  
The infrared behavior of the theory has a dual, 
magnetic description in terms of an $SO(2N_{f}-2N_{c}+4)$ gauge theory with
$2N_{f}$ flavors of dual quarks and an additional gauge singlet field
$M^{ij}$. A superpotential takes the form of 
\beq
W = \frac{1}{2\lambda}M^{ij}q_{i}q_{j}
\eeq
where $\lambda$ relates the scale of electric theory $\Lambda_{N=1}$ and those 
of the magnetic theory $\tilde{\Lambda}_{N=1}$ by 
\beq
\Lambda^{3(N_{c}-1)-N_{f}}\tilde{\Lambda}^{3(N_{f}-N_{c}+1)-N_{f}} =
C (-1)^{N_{f}-N_{c}}\mu^{N_{f}}
\eeq
Besides, there are also gauge invariant operators given in the electric theory by 
\bea
B^{[i_{1},\cdots, i_{2N_{c}}]} &=& Q^{i_{1}}\cdots Q^{i_{2N_{c}}} \\ \nonumber  
b^{[i_{1},\cdots, i_{2N_{c}-4}]} &=& W^{2}_{\alpha} Q^{i_{1}}\cdots 
Q^{i_{2N_{c}-4}} \\ \nonumber
{{\cal W}}_{\alpha}^{[i_{1},\cdots,i_{2N_{c}-2}]} &=& W_{\alpha}Q^{i_{1}}\cdots 
Q^{i_{2N_{c}-2}} \
\eea
Actually, the last two types of baryons, $b$ and $\cal W$, occur in the case of 
mixed Coulomb-Higgs branches. This may appear when some vev for $Q$'s are
not zero but not all of them are different from zero. Our values for the
vev of $M$ tells us that all of them can be zero (for Coulomb phase)
 or  be non-zero (for Higgs phase). So there is
no mixed Coulomb-Higgs branches. That is,  in our theory 
$b$ and $\cal W$ are 0 identically, so only baryons of type $B$ do exist.
\subsection{$SO(2N_{c}+1)$ case}

For $2N_{f}\le (2N_{c}+1)-5$,
we can continue to proceed with the arguments as in the case of 
$2N_{f}\le 2N_{c}-6$ and obtain the moduli space given by the vev of $M$.
For $2N_{f} = (2N_{c}+1)-3$, again there exist two branches, as explained in 
\cite{is}. The ADS superpotential is
\beq
W = 4 (1+\epsilon)\frac{\Lambda_{N=1}^{4N_{c}-1}}{\mbox{det M}}
\eeq
so we have again two physically inequivalent phase branches as a function 
of $\epsilon$. For $\epsilon=1$, this is  just a continuation 
of (\ref{eff}) to the case $2N_{f}=(2N_{c}+1)-3$, so there is  no
vacuum as $\mu$ goes to infinity.
For $\epsilon = -1$  there is no ADS superpotential. 
Again, what is to be extrematized is $\Delta W$ which gives as moduli
space solution only $m=0$. In order to match the 't Hooft anomaly conditions,
we need to have supplementary fields $q_{i}$ at the origin, $M=0$, which have mass
away from $M=0$. The most general invariant superpotential which describes the
coupling of $q$'s and $M$  can be written as  \cite{is}
\beq
W = \frac{1}{2\lambda}f\left(t=\frac{(\mbox{det} M)(M^{ij}q_{i}q_{j})}{
\Lambda_{N=1}^{2N_{c}
-2}}\right) M^{ij}q_{i}q_{j}
\eeq
where f(t) must be holomorphic near $t = 0$ and $f(0) = 1$.
The massless fields $q_{i}$ are identified with the ``exotic'' composites
$b_{i}$ as in \cite{is}, and they confine at $M=0$ giving again confinement
without chiral symmetry breaking.  
For $2N_{f}\ge 2N_{c}-1$ there is a dual description of the theory with
gauge group $SO(2N_{f}-2N_{c}-4)$, with dual quarks $q^{i}$ in the
fundamental of the magnetic group and with singlets $M^{ij}$. 

We close this section by noting
that for $SO$ group we do not have non-baryonic branches as compared 
with $SU$ group because of the specific form of the solution for $Q$'s
which solve the D-term and F-term equations \cite{aps}. This simplifies
our discussion  as compared with $SU$ group.

\section{$N=2$ Higgs Branch from M Theory }
\setcounter{equation}{0}

We study the moduli space of vacua of $N=2$ supersymmetric
$SO(N_c)$ gauge theory with $N_f$ flavors by analyzing M theory 
fivebrane. Let us first describe the Higgs branch in the type
IIA brane configuration and later go on to the M theory fivebrane in terms of
geometrical picture.

The brane
configuration consists of two parallel NS5
branes extending in the direction $(x^0, x^1, x^2, x^3, x^4, x^5)$, 
the D4 branes
stretched between two NS5 branes and extending 
over $(x^0, x^1, x^2, x^3)$ and
being finite in $x^6$ direction, and the D6 branes extending 
in the direction
of $(x^0, x^1, x^2, x^3, x^7, x^8, x^9)$. 
In order to get orthogonal gauge group, we should consider an O4
orientifold which is parallel to the D4 branes in order not to break the
supersymmetry and is infinite extent along the $x^6$ direction. 
The O4 orientifold
gives a spacetime reflection as $(x^4, x^5, x^7, x^8, x^9) \rightarrow
(-x^4, -x^5, -x^7, -x^8, -x^9)$, in addition to the gauging of worldsheet
parity $\Omega$.
Each object which does not lie at the fixed points of the spacetime symmetry
( i.e., over the O4 orientifold ) must have its mirror image. 
Thus NS5 branes have a mirror in $(x^4,
x^5)$ directions and D6 branes have a mirror in $(x^7, x^8, x^9)$ directions.
For $SO(2N_c)$ gauge group, each D4 brane at $v=x^4+i x^5$
has its mirror image at $-v$: $N_c$ D4 branes and its mirror $N_c$ ones.
Similarly, for the $SO(2N_c+1)$ gauge group, there exist an extra single
D4 brane which lies over the O4 orientifold and is frozen at $v=0$ because
it does not contain its mirror image, as well as 
$N_c$ D4 branes and its mirror $N_c$ ones.
Another important ingredient of O4 orientifold is its charge, which is
related to the sign of $\Omega^2$. When the
D4 brane carries one unit of this charge, the charge of the O4 orientifold 
is $\pm 1$, for $\Omega^2= \mp 1$ in the D4  brane sector.

To go to the Higgs branch, the D4 branes
are broken on the D6 branes and suspended between those D6 branes
being allowed to move on the directions
$(x^{7}, x^{8}, x^{9})$. 
The dimension of the Higgs moduli space is obtained 
by counting all possible breakings of D4 branes into D6 branes 
as follows: the first D4 brane is broken in
$N_{f}-1$ sectors between the D6 branes, the second D4 branes is
broken in $N_{f}-3$ sectors 
and so on. Besides that we have to consider the symmetric orientifold
projection which increases degrees of freedom, as explained in
\cite{egkrs}. 
Then the dimension of the Higgs moduli space is given by for $SO(N_c)$:
\beq
\label{higgsdim}
2[(2N_{f}-2+1) + (2N_{f}-6+1) + \cdots + (2N_{f}-2N_c+2+1)]=
N_c(2N_f-N_c +1)
\eeq
or $2N_cN_{f} - N_c(N_c-1)$ where we have
explicitly added 1 as a result of the symmetric orientifold 
projection. The overall factor $2$ in the left hand side is due 
to the mirror D6 branes and the result is 
very similar to the field theory result where because of the $N_{f}$ vevs, 
the gauge symmetry is completely broken and there are 
$N_cN_{f}-\frac{1}{2}N_c(N_c-1)$ massless neutral hypermultiplets. 
Thus, the field theory results match
the brane configuration results.
There is no intersection
between the Higgs and Coulomb branches and we have only 
{\it one Higgs branch}.

Let us describe how the above brane configuration is embedded in 
M theory in terms of
a single M fivebrane whose worldvolume is
${\bf R^{1,3}} \times \Sigma$, where $\Sigma$ is identified with Seiberg-Witten
curves that determine  the solutions to Coulomb branch 
of the field theory.
As usual, we write $s=(x^6+i x^{10})/R, t=e^{-s}$,
where $x^{10}$ is the eleventh coordinate and is compactified
on a circle of radius $R$. Then the curve $\Sigma$, describing
 $N=2$  $SO(N_c)$ gauge theory with $N_f$ flavors and even $N_c$,
is given\cite{lll} by an equation in $(v, t)$ space
\bea
\label{soeven}
v^2 t^2-B(v^2, u_k )t+
\Lambda_{N=2}^{2N_c-4-2N_f} v^2 \prod_{i=1}^{N_f} (v^2-{m_i}^2)=0.
\eea
Here  $B(v^2, u_k)$ is a degree $N_c$ polynomial of the form 
$v^{N_c} + u_2 v^{N_c -2} + \cdots + u_{N_c}$ with only even degree terms
and  the coefficients
depending on the moduli $u_k$, and $m_i$ is the mass of quark.
Similarly, for odd $N_c$, 
it takes the form of
\bea
\label{soodd}
v t^2-B(v^2, u_k )t+
\Lambda_{N=2}^{2N_c-4-2N_f} v \prod_{i=1}^{N_f} (v^2-{m_i}^2)=0
\eea
where this time $B(v^{2})$ is a polynomial of degree $N_{c}-1$.

\subsection{Including D6 Branes }

In M theory, the type IIA D6 branes are the magnetic dual of the 
electrically charged  D0 branes, 
which  are the Kaluza-Klein monopoles described by a Taub-NUT space. 
We will ignore the
hyper-K\"ahler structure of this Taub-NUT space and instead use one  of the 
complex structures, which
can be  described by \cite{lll} 
\bea
\label{d6}
y z=\Lambda_{N=2}^{2N_c-4-2N_f} \prod_{i=1}^{N_f} (v^2-{m_i}^2)
\eea
in $\bf C^3$ for $SO(N_c)$. 
The D6 branes are located at $y=z=0, v=\pm m_i$.
This surface, which represents a nontrivial $\bf S^1$ bundle over 
$\bf R^3$ instead of
the flat four dimensional space ${\bf R^3} \times {\bf{S^1}}$ with coordinates
 $(x^4, x^5, x^6, x^{10})$,
is the unfolding of the $A_{2n-1}$ ($n=N_f$) singularity in general. 
The Riemann surface $\Sigma$ is embedded as a curve in this curved surface and
given by
\bea
\label{sigmaeven}
v^2(y+z)= B(v^2)
\eea
when  $N_c$ is even.
In the case $N_c$ is odd, the curve is given by
\bea
\label{sigmaodd}
v(y+z) = B(v^2).
\eea
Our type IIA brane
configuration has $U(1)_{4,5}$ and $SU(2)_{7,8,9}$ symmetries regarded as
classical $U(1)$ and $SU(2)$ R-symmetry groups of the 4 dimensional theory
on the brane worldvolume. One of them, 
$SU(2)_{7,8,9}$, is preserved only in M theory quantum 
mechanical configuration, but
$U(1)_{4,5}$ is broken. The $U(1)_{R}$ 
symmetry of the $N=2$ supersymmetric field theory is anomalous being broken by
instantons whose factor is proportional 
with $\Lambda_{N=2}^{2N_c-4-2N_{f}}$. 
We can easily see that the charge of $\Lambda_{N=2}^{2N_c-4-2N_{f}}$
is $(4N_c-8-2N_{f})$ from equations (\ref{d6}) by considering $v$ of charge 2. 

\subsection{Resolution of Singularities and the Higgs Branch}

In this section, we follow the notations of \cite{aot1} and refer to
\cite{aot1} for details.
When all the bare masses are the same but not zero (say $m=m_i$), the surface
(\ref{d6}) develops $A_{N_f -1}$ singularities at two points
$y =z =0, v=\pm m$. Over each singular point, there are $(N_f-1)$
rational curves on the resolved surface. We denote the rational curves
over the point $y=z=0, v=m$ by $C_1, C_2, \cdots C_{N_f -1}$ and those over
the point $y=z=0, v= -m$ by $C'_1, C'_2, \cdots C'_{N_f -1}$.
When we turn off the bare mass, i.e., $m_i =0$ for all $i$, the singularity
gets worse and  a new rational curve appears in the resolved surface which 
we call $E$.

Now, we would like to study the Higgs branch when all the bare masses are
 turned
off.
As noticed in \cite{aot1,hoo}, the complex structure  of Taub-NUT space
develops $A$-type singularities when all bare masses become massless.
In M theory,  the Higgs branch appears when the fivebrane
intersects with the D6-branes. As a special case, we will consider
  when all D4 branes are broken on $D6$ branes in the type IIA picture.
To describe the Higgs branch, we will study 
how the curve 
\bea
y+z = v^{N_c}
\eea
appears in the resolved $A_{2N_f -1}$ surface since there are
 $N_c$ D4 branes in this configuration.
Away from the singular point $y=z=v=0$, we may regard the curve
 as embedded in the original
 $y-z-v$ space because there is no change in the resolved surface in this region.
 Near the singular point $y=z=v=0$, we have to consider the resolved surface.  
On the $i$-th patch  $U_i$ of the resolved
surface, the equation  of the curve $\Sigma$ becomes
\bea
y^i_i z^{i-1}_i +  y_i^{2N_f-i}z_i^{2N_f+1-i} =  y_i^{N_c}z_i^{N_c}
\eea
By factorizing this equation according to the range of $i$, we will see that
the curve consists of several components. One component, 
which we call $C$, is the extension of the one in the region away from 
$y=z=v=0$ which
we have already considered. The other components are the rational curves $C_1, 
\ldots , C_{n-1}, E,
C'_1,\ldots , C'_{n-1}$ with some multiplicities. For convenience, we rename the exceptional
curves $E_1, \ldots , E_{2n-1}$ so that $E_i$ is defined by $y_i=0$ on $U_i$ and 
$z_{i+1} =0$ on $U_{i+1}$.
Hence we can see from the above factorization that the component $E_i$ has  
multiplicity $l_i = i$ for 
$i=1,\ldots , N_c $; $l_i =N_c $ for $i=N_c +1, \ldots , 2N_f -N_c $; and 
$l_i = {2N_f-i}$ for
$i=2 N_f- N_c +1,\ldots  2N_f-1$.   
 Note that the component $C$ intersects with $E_{N_c}$ and $E_{2N_f -N_c}$.

To count the dimension of the Higgs branch, recall
that once the curve degenerates and $\bf CP^1$ components are generated, they can move
in the $x^7, x^8, x^9$ directions \cite{w1}.
 This motion together with the integration of the chiral two-forms on such
$\bf CP^1$'s parameterizes the Higgs branch of the four-dimensional theory.
While the components $C_i$ and $C'_i$ have to move in pairs due to orientifolding,
the middle component $E_{N_f} =E$ moves freely because it is mirror symmetric
with respect to O4 plane. 
Thus  the quaternionic dimension of the Higgs branch is 
\bea
\frac{1}{2}(\sum_{i=1}^{2N_f-1} l_i + l_{N_f})= \sum_{i=1}^{N_c-1} i + (2N_f-2N_c +2 )N_c=
\frac{1}{2}N_c(2N_f-N_c +1),
\eea
which is the half of the complex dimension given in (\ref{higgsdim}). 

\section{$N=1$ SQCD}
\subsection{The Rotated Configuration}
\setcounter{equation}{0}

By adding a mass term to the adjoint chiral multiplet, $N=2$ SUSY will
be broken to $N=1$. 
To describe the corresponding brane configuration in M theory,
we introduce a complex coordinate
\bea
w=x^8+i x^9.
\eea
Before breaking the $N=2$ supersymmetry, the fivebrane is located at $w=0$.
Now we rotate only the left NS5 brane toward the $w$ direction
 and from the behavior of
two asymptotic regions which correspond to the NS5(left) and NS'5(right) 
branes with 
$v \rightarrow 
\infty$ this rotation leads to the following boundary conditions 
for $SO(N_c)$:
\bea 
\label{bdy-cond}
& & w \rightarrow \mu v \;\;\; \mbox{as}\;\;\; v \rightarrow 
\infty, \;\;\; t \sim v^{
N_c-2}  \nonumber \\
& & w \rightarrow 0 \;\;\; \mbox{as}\; \;\; v \rightarrow 
\infty, \;\;\; t \sim
\Lambda_{N=2}^{2(N_c-2-N_f)}v^{2N_f-N_c+2}.
\eea
After
rotation, $SU(2)_{7,8,9}$ is broken to $U(1)_{8,9}$.
In order for this to be consistent, 
because of the relation between $v$ and $w$ in
(\ref{bdy-cond}), $\mu$ 
should have charges under $U(1)_{4,5}\times U(1)_{8,9}$. That is,
$v$ has charge 2
under $U(1)_{4,5}$ and 0 under $U(1)_{8,9}$ while $w$ has charge 0 
under $U(1)_{4,5}$
and 2 under $U(1)_{8,9}$.

Since the rotation is only possible at points in moduli space 
at which all 1-cycles 
on the curve $\Sigma$
are degenerate \cite{sw}, the curve $\Sigma$ is rational, which means that the
 functions $v$ and $t$ can
be expressed as some  rational functions of $w$ 
after we identify $\Sigma$ with a 
complex plane $w$ having
some deleted points. 
Because of the symmetry of $v \rightarrow -v$ and $ w \rightarrow -w$ due to 
orientifolding, we can write them as:
\bea
v^2=P(w^2), \;\;\;\;\; t=Q(w).
\eea
(Of course, we may write $t = Q(w^2)$ for even $N_c$
 and $t^2 = Q(w^2)$ for odd $N_c$ considering  the  extra
 symmetry $t \rightarrow -t$. But we  suppressed these extra
symmetries to treat both cases uniformly.)
Since $v$ and $t$ become $\infty$ only if $w=0$ and  $\infty$ from the
boundary conditions, these 
rational functions are 
some polynomials of $w$ up to
a factor of some power of $w$: $P(w^2) = w^{2a}p(w^2)$ and $Q(w) = 
w^{b}q(w)$ where 
$a$ and $b$ are some 
integers and $p(w^2)$ and $q(w)$ are some polynomials of 
$w$ which we may assume 
non-vanishing at $w=0$. 
Near one of the points at $w=\infty$, 
$v$ and $t$ behave 
as $v\sim \mu^{-1}w$ and
$t \sim v^{N_c -2}$ by (\ref{bdy-cond}). Thus the rational 
functions are of the form
\bea
P(w^2) = w^{2a}(w^{2-2a} + \cdots)/\mu^{2}\;\;\; \mbox{and}\;\;\; 
Q(w) = \mu^{-N_c +2}w^{b}(w^{N_c -2-b} + \cdots)
\eea
for $SO(N_c)$.  
Around a neighborhood $w=0$, 
the Riemann surface $\Sigma$ can be parameterized 
by $1/v$ which vanishes as 
$w\to 0$ from the boundary condition.
Since $w$ and $1/v$ are two coordinates around the neighborhood $w=0$ in the 
compactification of $\Sigma$
and vanish at the same point, they must be linearly proportional to
each other 
$w \sim 1/v$ in the 
limit $w \to 0$.
The numerator $w^{2-2a} + \cdots$ of  $P(w^2)$ then takes the form 
$(w^2 - w_+)(w^2 -w_-)$. But
the equation of $v^2 = P(w^2)$ implies 
that $P(w^2)$ must be a square of a rational function. Hence we have
$w_+ = w_-$ and by letting $w_0^2 = w_\pm$
\bea
\label{rot2}
v^2=P(w^2)=\frac{(w^2-w_0^2)^2}{\mu^2 w^2}.
\eea
Since $t\sim  v^{2N_f -N_c +2}$ and $w\sim 1/v$ as $w\to 0$, we get 
$b= N_c -2 -2N_f$ and thus,
\bea
t=Q(w) = \mu^{-N_c +2}w^{N_c-2 -2N_f}(w^{2N_f} + \cdots ).
\eea
By the equation $yz = v^{2N_f}$ where $N_f > 0$ defining  the space-time, 
$t=0$ (i.e., $y=0$) implies
$v=0$. Therefore the zeros of the polynomial $w^{2N_f} + \cdots$ are equal to
the zeros $\pm w_0$ of $P(w^2)$. Hence we conclude
\bea
\label{rot3}
t=Q(w)=\mu^{-N_c+2} w^{N_c-2-2N_f} (w^2-w_0^2)^{N_f}.
\eea
Near $w =\pm  w_0$, $v$ and $t$ will satisfy
the relation (\ref{soeven}) (resp. (\ref{soodd}))
 for even (resp. odd) $N_c$ 
 which is approximated in the limit $v\to \infty$
  by
\bea
t^2 - v^{N_c -2}t +\Lambda_{N=2}^{2N_c-4-2N_f} v^{2N_f} = 0
\eea
By plugging the equations (\ref{rot2}) and (\ref{rot3}) in the
this equation and considering the limit $w \to \pm w_0$,
 we obtain the expression for $w_0$:
\begin{equation}
\label{w3}
(-1)^{N_{f}} \Lambda_{N=2}^{2N_c-4-2N_f}
w_{0}^{2N_{f}}\mu^{2N_c-4-2N_{f}} = w_{0}^{2N_c-4}
\end{equation}
This yields a first solution 
\begin{equation}
\label{w4}
w_{0} = (-1)^{\frac{N_f}{2N_c-4-2N_f}} \mu \Lambda_{N=2}.
\end{equation}
and a second possible solution
\begin{equation}
\label{w5}
w_{0} = 0
\end{equation}

As we have noticed in section 2, 
if $2N_f=N_c-4$ (resp. $2N_f = N_c -3$)
for even (resp. odd) $N_c$, the global chiral symmetry is not broken
 at the origin so there exists a confinement at the origin 
without chiral symmetry breaking.
Now, by sending the D6 branes to infinity in the $x_{6}$ direction, there
are $N_{f}$ semi-infinite D4 branes ending on the right NS5 brane from the
right, whose asymptotic positions in the $w$ direction are just given by
the eigenvalues of the meson matrix $M$. In the $Sp$ case, the solution (\ref{w4})
was the only one taken in \cite{aot1}. 
 The solution of (\ref{w5}) corresponds to a zero eigenvalue of
$M$ (the origin) and  the chiral symmetry is broken at the origin
for $Sp$ case so  $w_{0}=0$ is not an acceptable solution. 
However, in the case
of $SO$ groups with $2N_{f}=N_c-4$ or  $2N_{f}=N_c -3$, 
we have confinement without chiral symmetry breaking at the origin so
$w_{0}$ becomes an acceptable solution. From the RG equation, we observe that,
for $2N_{f}\ge 2N_{c}$, the solution (\ref{w4}) goes to zero also so the two
values for $w_{0}$ coincide. For $2N_{f}=2N_{c}-2,2N_{c}-3,2N_{c}-4$,
both solutions for $w_{0}$ exist and they give the two branches that we have discussed
in section 2. For $2N_{f}\le 2N_{c}-5$, again using the RG equation, we observe
that in the left hand side of (\ref{w3}) the power of $\mu$ becomes positive so
it will
become infinite in the $\mu\rightarrow\infty$ limit. Therefore the equation
(\ref{w3}) cannot have the solution of (\ref{w5}) because in the left hand side
we will have the product of zero with $\infty$ which is undetermined.
So, for $2N_{f}\le 2N_{c}-5$, there is only one solution given by (\ref{w4})
which becomes infinite in this range so, as expected, there is no vacuum.

\subsection{ $SO(2N_{c})$ with  massless matter}

After rescaling $t$ by a factor $\mu^{2N_c -2}$, 
when we consider the case of $2N_{f}$ massless quarks,
 the rotated curves are described by :
\bea
\label{curve}
\mu^{-(2N_c -2)}v^2 \tilde{t}^2 &-& B(v^2, u_k) \tilde{t} \,\, +\,\,
 \Lambda_{N=1}^{3(2N_{c}-2)-2N_{f}} v^{2+2N_f}\,\, =\,\, 0 \\  \nonumber
\tilde{t} &=& w^{2(N_{c}-1-N_{f})} (w^{2}-w_{0}^{2})^{N_{f}} \\  \nonumber
v w       &=& \mu^{-1} (w^{2}-w_{0}^{2})\
\eea 
where $\tilde{t} = \mu^{2N_c -2}t$.
Since the order parameters $u_k$ vanish in the $\mu \to \infty$ limit,
the first equation has the smooth limit given by
\bea
\tilde{t} = \Lambda_{N=1}^{3(2N_{c}-2)-2N_{f}} v^{2N_{f}-(2N_{c}-2)}.
\eea
By the RG equation
\beq
\Lambda_{N=1}^{3(2N_{c}-2)-2N_{f}} = \mu^{2N_{c}-2}
\Lambda_{N=2}^{2(2N_{c}-2)-2N_{f}}.
\eeq
we may classify the rotated branes into  several cases:

$\bullet \;\;\;2N_{f} \ge 2N_{c}\qquad$
The $\mu\Lambda_{N=2}$ goes to zero for $\mu\rightarrow\infty$ limit. Thus
there is no difference between the two possible values for $w_{0}$ 
we have discussed before.

$\bullet \;\;\;2N_{f} \le 2N_{c}-6\qquad$
The $\mu\Lambda_{N=2}$ goes to infinity so there is no supersymmetric
vacua.

$\bullet \;\;\;2N_{f} = 2N_{c}-4\qquad$
For the value $w_{0}=\mu\Lambda_{N=2}$ we again do not have
any vacuum, this case being just a continuation of the previous case. On the other hand,
for $w_{0}=0$ there exists  a vacuum at the origin, and the curve is given by:
\bea
v = 0\qquad
\tilde{t}& =& w^{2N_c -2}
\eea

$\bullet \;\;\;2N_{f} = 2N_{c}-2\qquad$
We have two possible curves, one for $w_{0}=0$, the
other for $w_{0}=(-1)^{-N_f/2}\mu\Lambda_{N=2}$.
We obtain as in 
\cite{hoo} the left component $C_{L}$ and the
right component $C_{R}$: 
\bea
C_{L} \left\{ \begin{array}{lcl} 
v& =& 0\\
\tilde{t} &=& (w^2 - w_0^2)^{N_c - 1}
\end{array}
\right.
\qquad
C_{R}\left\{ \begin{array}{lcl}
w &= & 0\\
\tilde{t}& =& \Lambda_{N=1}^{4N_{c}-4}
\end{array}
\right.
\eea
Of course, $C_{L}$
differs as a function of the value of $w_{0}$.
Furthermore, two components  $C_L$ and $C_R$ intersect at the point $v=w=0$,
$\tilde{t} = \Lambda_{N=1}^{4N_{c}-4}$ for
$w_{0}=(-1)^{-N_f/2}\mu\Lambda_{N=2}$ whereas $C_L$ does not meet
$C_R$ for $w_0 = 0$.

\subsection{$SO(2N_{c}+1)$ with  massless matter}

By similar reasoning, the rotated brane for the case of $2N_{f}$ massless quarks 
is described by:
\bea
\tilde{t}^{2} &=& w^{4N_{c}-2-4N_{f}}(w^{2}-w_{0}^{2})^{2N_{f}} \\ \nonumber
 v w      &=& \mu^{-1}(w^{2}-w_{0}^{2}) \
\eea 
Again we have several cases, arising from the RG equation
\beq
\Lambda_{N=1}^{3(2N_{c}-1)-2N_{f}} = 
\mu^{2N_{c}-1}\Lambda_{N=2}^{2(2N_{c}-1)-2N_{f}}:
\eeq

$\bullet \;\;\; 2N_{f}\ge 2N_{c}+1\qquad$
The  $\mu\Lambda_{N=2}$ vanishes again  for $\mu\rightarrow\infty$ limit.
Thus $w_{0}=\mu\Lambda_{N=2}$ goes to zero and the two branches unify for this
values of $N_{f}$.

$\bullet \;\;\; 2N_{f}\le 2N_{c}-5\qquad$
The $\mu\Lambda_{N=2}$ goes to infinity so there is no supersymmetric
vacuum.

$\bullet \;\;\; 2N_{f}=2N_{c}-3\qquad$
For $w_{0}=\mu\Lambda_{N=2}$, this again goes to infinity, and there is no
vacuum. For $w_{0}=0$ there is a vacuum at the origin, where there are additional
massless field, besides the meson. The curve describing the vacuum is
\bea
v = 0 \qquad
\tilde{t}^{2} &=& w^{4N_{c}-2}
\eea

$\bullet \;\;\; 2N_{f}=2N_{c}-1\qquad$
the $\mu\Lambda_{N=2}$ does not depend on $\mu$ so both solutions
for $w_{0}$ are possible. Again $C_{L}$ depends on the value of $w_{0}$.
Even if this case resembles the case of $SU(N_{c})$ for $N_{f}=N_{c}$
and $SO(2N_{c})$ for $2N_{f}=2N_{c}-2$ in the sense that $\mu\Lambda_{N=2}$
does not depend on $\mu$, here the difference is that the dual is a non
Abelian theory so we do not have effects like the appearance of a supplementary
superpotential (as in $SU(N_{c})$ case) or the appearance of monopoles and dyons
at vacua (as in $SO(2N_{c})$ case).

This completes the discussion for the rotated brane in the presence of
massless matter. For discussions on dualities for theories
 with $SO$ groups from 
M theory we refer to \cite{cs}.

\section{Conclusion}
\setcounter{equation}{0}

In the present work, we generalized the work of \cite{w2,hoo,bo,aot1}
to the case of $SO(2N_{c})$ and $SO(2N_{c}+1)$ gauge groups. We showed that the
brane configuration gave us information about the $N=1$ field
theory vacua. We discussed the rotation of the M fivebrane from $N=2$ theory
to $N=1$ theory and we obtained the form of the rotated curves. We want to
emphasize again that we were not able to obtain the spectrum of massless 
particles at different vacua either in field theory or in the brane configuration
picture. The same problem appeared in \cite{aps} and seems to be a common
flaw of the $N=1$ theories obtained from $N=2$ by turning mass for the
adjoint (in field theories) or by rotating the branes (in the brane configuration
picture). A possible way to obtain information about the massless particles
is to use the idea developed very recently by Strassler in \cite{str}. In the context
of field theory, he found that the electric sources in the spinor 
representations can be introduced as magnetic sources in the dual 
nonabelian gauge theory. By turning masses for some quarks one can thus obtain
information about the spectrum of particles at vacua for theories with smaller
number of quarks like the theories with $N_{f}=N_{c}-2$ or $N_{f}=N_{c}-4$. 
A first step towards this direction would be to
generalize the construction of \cite{lyk} for the $SO$ gauge group.
We hope that this problem will be solved in the near future.

\end{document}